\documentclass[12pt,preprint]{aastex}

\newcommand{\eq}{\begin{equation}}
\newcommand{\ee}{\end{equation}}

\def\t0{\theta_{\circ}}

\def\be{\begin{equation}}
\def\en{\end{equation}}

\def\gapp{\ \lower 3pt\hbox{${\buildrel > \over \sim}$}\ }
\def\lapp{\ \lower 3pt\hbox{${\buildrel < \over \sim}$}\ }
\begin{document}

\title{Multi-color Shallow Decay and Chromatic Breaks in the
GRB\,050319 Optical Afterglow}
 
\author{
K.Y. \textsc{Huang}\altaffilmark{1}, 
Y. \textsc{Urata}\altaffilmark{2,3},
P.H. \textsc{Kuo}\altaffilmark{1},
W.H. \textsc{Ip}\altaffilmark{1},
K. \textsc{Ioka}\altaffilmark{4}, 
T. \textsc{Aoki}\altaffilmark{5}, 
C.W. \textsc{Chen}\altaffilmark{1}, 
W.P. \textsc{Chen}\altaffilmark{1},
M. \textsc{Isogai}\altaffilmark{5},
H.C. \textsc{Lin}\altaffilmark{1},
K. \textsc{Makishima}\altaffilmark{3,6},
H. \textsc{Mito}\altaffilmark{5},
T. \textsc{Miyata}\altaffilmark{5},
Y. \textsc{Nakada}\altaffilmark{5},
S. \textsc{Nishiura}\altaffilmark{7},
K. \textsc{Onda}\altaffilmark{2},
Y. \textsc{Qiu}\altaffilmark{8},
T. \textsc{Soyano}\altaffilmark{5},
T. \textsc{Tamagawa}\altaffilmark{3}, 
K. \textsc{Tarusawa}\altaffilmark{5},
M. \textsc{Tashiro}\altaffilmark{2},
and
T. \textsc{Yoshioka}\altaffilmark{9}
}

\altaffiltext{1}{Institute of Astronomy, National Central University, Chung-Li 32054, Taiwan, Republic of China. (Huang email: d919003@astro.ncu.edu.tw)}
\altaffiltext{2}{Department of Physics, Saitama University, Shimo-Okubo, Sakura, Saitama, 338-8570, Japan.}
\altaffiltext{3}{RIKEN, 2-1 Hirosawa, Wako, Saitama 351-0198, Japan.}
\altaffiltext{4}{Department of Physics, Kyoto University, Kyoto 606-8602, Japan}
\altaffiltext{5}{Kiso Observatory, Institute of Astronomy, The University of Tokyo, Mitake-mura, Kiso-gun, Nagano 397-0101, Japan.}
\altaffiltext{6}{Department of Physics, University of Tokyo, 7-3-1 Hongo, Bunkyo-ku, Tokyo 113-0033, Japan.} 
\altaffiltext{7}{Department of Astronomy and Earth Sciences, Tokyo Gakugei University, Koganei, Tokyo 184-8501, Japan.}
\altaffiltext{8}{National Astronomical Observatories, Chinese Academy of Sciences, Beijing 100012, China, PR.}            
\altaffiltext{9}{Department of Physics, Nagoya University, Furo-cho, Chikusa, Nagoya, Aichi 464-8602, Japan.}

%\authoremail
 
\begin{abstract}

Multi-wavelength $B$, $V$, $R$, $I$ observations of the optical
afterglow of GRB\,050319 were performed by the 1.05-m telescope at
Kiso Observatory and the 1.0-m telescope at Lulin Observatory from
1.31 hours to 9.92 hours after the burst. Our $R$ band lightcurves,
combined with other published data, can be described by the smooth
broken power-law function, with $\alpha_1$ = $-$0.84 $\pm$0.02 to
$\alpha_2$ = $-$0.48$\pm$0.03, 0.04 days after the GRB. The optical
lightcurves are characterized by shallow decays--- as was also
observed in the X-rays--- which may have a similar origin, related to
energy injection. However, our observations indicate that there is
still a puzzle concerning the chromatic breaks in the $R$ band
lightcurve (at 0.04~days) and the X-ray lightcurve (at 0.004~days)
that remains to be solved.
\end{abstract}
 
 
\section{Introduction}

  The GRB afterglow as perceived in the X-ray, optical and radio
wavelengths is now understood to be the result of the collision
between relativistic ejecta from the gamma-ray bursts and the
interstellar medium (ISM). A comparison of afterglow lightcurves
obtained at different wavelengths gives important information about
the surrounding ISM environment and the interaction processes. Such
analyses can also provide essential input for theoretical
models. Recently, the pace of this type of activity has quickened
significantly, stimulated by the capabilities of the quick response
and accurate localization of the GRB by the {\it Swift} Satellite
(Gehrels et al. 2004). This has meant that the number of GRB optical
afterglow detections in the first several hours by ground-based
telescopes has recently increased significantly.  It is interesting to
note that the observations by {\it Swift} of the early X-ray emissions
from a number of GRBs reveal a canonical behavior. The X-ray
lightcurves can be divided into three distinct power law segments
(Nousek et al. 2006). Some X-ray and optical observations show that
the evolution of both lightcurves changes at the same time (Blustin et
al. 2006; Rykoff et al. 2006), however chromatic breaks were also
found in some cases (Fan \& Piran 2006; Panaitescu et al. 2006). The
nature of the afterglow early breaks in the lightcurves is thus
uncertain. A detailed comparison of changes in the evolution of the
optical, radio and X-ray lightcurves should therefore be very
interesting.  This kind of physical study demands both a
well-coordinated observational program and careful data analysis. We
use GRB\,050319 which has comprehensive observational coverage in both
the X-ray and optical wavelengths may be used as just such an example.

  GRB\,050319 was detected by the Burst Alert Telescope (BAT)
instrument onboard the {\it Swift} satellite on March 19, 2005 at
09:31:18.44 UT (Krimm et al. 2005). However, a re-analysis of the BAT
data showed two flares, which indicated that GRB\,050319 had already
started 137 sec before the trigger. The 15--350~keV fluence for the
entire burst duration of $T_{90}= 149.6\pm0.7$ sec has been estimated
to be 1.6 $\times 10^{-6}~{\rm erg\, cm}^{-2}$ (Cusumano et
al. 2006). The X-ray emission of GRB\,050319 after the burst was
monitored by the XRT from 225 sec to 28 days\footnote[1] { The burst
time in the article is 09:29:01.44 UT, 137~s before the BAT
trigger}. Two breaks in the emission curves were found (Cusumano et
al. 2006). The initial sharp decline can be described by a power-law
with an index of $\alpha_{1}= -5.53\pm 0.67$ to be followed by
$\alpha_{2}= -0.54\pm 0.04$ after 0.004 days since the burst. The
unusually flat decline in the second part might have been caused by
continuous energy injection (Cusumano et al. 2006). At about 0.313
days after the burst, the power-law index changed to $\alpha_{3}=
-1.14\pm 0.2$ which can be readily explained as a jet break or a
reduction in the energy injection (Cusumano et al. 2006; Zhang et
al. 2006) .

The early optical afterglow emission at 230~s was observed by the UVOT
telescope on {\it Swift} (Mason et al. 2006) and by two ground-based
robotic telescopes, {\it ROTSE-III} (Quimby et al., 2006) and {\it
RAPTOR} (Wo\'zniak et al. 2005). The best single power-law fit of
unfiltered data from {\it ROTSE-III} and {\it RAPTOR} indicates that
$\alpha = -0.854\pm 0.014$. A number of optical observatories have
joined the follow-up observations (see Yoshioka et al. 2005; Torii et
al. 2005; Sharapov et al. 2005a,b; George et al. 2006; Misra et
al. 2005; Kiziloglu et al. 2005; Greco et al. 2005). The spectral
measurements of the afterglow by the Nordic Optical Telescope (NOT)
indicate that was redshift $z=3.24$ of this event (Fynbo et al. 2005).

\section{Observations and Analysis}

  After receiving the GRB alert message from {\it Swift} and the
afterglow position was reported by Rykoff et al. (2005), the
Target-of-Opportunity procedures of the East-Asia GRB Follow-up
Observation Network (EAFON\footnote[2]{EAFON web-page:
http://cosmic.riken.jp/grb/eafon/}, Urata et al. 2005) were
immediately carried out. A series of multi-band follow-up observations
were successfully performed by the 1.05-m Schmidt telescope of the
Kiso Observatory in Japan and the Lulin One-meter Telescope (LOT) in
Taiwan. Photometric $B$ and $R$ images were obtained at the Kiso site
with a 2k $\times$ 2k CCD camera (Urata et al. 2005) between 0.055 and
0.326 days after the burst. A number of parallel $B$, $V$, $R$ and $I$
images were obtained by LOT with a PI1300 CCD camera (Kinoshita et
al. 2005) from 0.080 to 0.413 days after the burst.

  The standard routine including bias subtraction and dark
subtraction; flat-field corrections were employed with the appropriate
calibration data needed to process the data using IRAF. The afterglow
can be clearly seen in the images. The signal-to-noise ratio was
improved by combing the LOT $B$ band data with median filtering. The
DAOPHOT package (Stetson 1989) was then used to perform point-spread
function (PSF) fitting for the GRB images. Four field stars were used
to create a PSF model which was applied to the optical afterglow of
each GRB image. For absolute photometric calibration, we used
calibrated data of the GRB field obtained by the USNOFS 1.0-m
telescope (Henden 2005). The photometric error and the systematic
calibration error were included in the magnitude error
estimation\footnote[3]{The errors in this article were quoted for
68$\%$ (1-$\sigma$) confidence level}.
\section{Results}
  
\subsection{Lightcurve}
 Figure\,1 shows the multi-band lightcurves of the GRB\,050319
afterglow.  Besides our $B$, $V$, $R$ and $I$ band data (Table\,1) we
also included the $R$ band measurements from {\it ROTSE-III} (Quimby
et al. 2006), {\it RAPTOR} (Wo\'zniak et al. 2005) and several GCN
reports (Greco et al. 2005; Kiziloglu et al. 2005; Misra et al. 2005;
Sharapov et al. 2005a,b). In addition, we also made use of several $B$
and $V$ band measurements taken by the {\it Swift} UVOT (Mason et
al. 2006). The GCN $R$ band points were re-calibrated using the
GRB\,050319 field stars reported by Henden (2005) so they could be
plotted on the same magnitude scale. The magnitude differences between
photometric field stars in Henden (2005) and in USNO-A2.0, USNO-B1.0
stars are $+0.18$ and $-0.22$ mag, respectively. We re-measured the
reference stars from Greco et al. (2005) from the LOT $R$ band images
and obtained the average magnitudes and rms errors.

After fitting the $B$, $V$, $R$ and $I$ band lightcurves to a single
power-law expression $F \propto t^{\alpha}$, where $\alpha$ is the
index and $t$ is the time after the burst, we get $\alpha = -0.56 \pm
0.06$ ($\chi^{2}/\nu$= 2.90 for $\nu=19$) for the $B$ band, $\alpha =
-0.65 \pm 0.03$ ($\chi^{2}/\nu$= 2.60 for $\nu=27$) for the $V$ band,
$\alpha = -0.59 \pm 0.01$ ($\chi^{2}/\nu$= 5.3 for $\nu=97$) for the
$R$ band, and $\alpha = -0.52 \pm 0.15$ ($\chi^{2}/\nu$= 7.7 for
$\nu=9$) for the $I$ band. This single power-law fitting indicates
that these lightcurves, obtained with different filters have similar
power-law decay even though the reduced chi-square values are
relatively large.

Since the data sets of the $V$ and $R$ measurements are more complete,
it is possible with following expression to attempt the fitting of the
corresponding lightcurves with a smoothly broken power-law function:

\begin{equation}
 F(\nu,t) = {2^{1/k}F_{\nu,b} \over [(t/t_{\rm b})^{-k\alpha_1}+(t/t_{\rm
 b})^{-k\alpha_2}]^{(1/k)}},
\end{equation}
where $t_{\rm b}$ is the break time, $\alpha_1$ and $\alpha_2$ are the
power-law indices before and after $t_{\rm b}$, $F_{\nu,b}$ is flux at
break $t_{\rm b}$, and $k$ is a smoothness factor. For the $V$ band,
we obtain $\alpha_1 = -0.87 \pm 0.21$, $\alpha_2 = -0.49 \pm 0.05$,
$t_{\rm b}$ = 0.042 $\pm$ 0.058 days and $k= -30$ ($\chi^{2}/\nu$=
1.48 for $\nu=24$). For the $R$ band, we obtain $\alpha_1 = -0.84 \pm
0.02$, $\alpha_2 = -0.48 \pm 0.03$, $t_{\rm b} = 0.046 \pm 0.008$
days, and $k = -21$ ($\chi^{2}/\nu$= 2.24 for $\nu=90$). This result
implies a mild break in both the $V$ and the $R$ band lightcurves at
around 0.04 days after the occurrence of the GRB.

Taking $t_{\rm b}$ = 0.04 days, we fit the data in the $B$ and $I$
bands to a respective power-law before and after the break. In this
manner, we find $\alpha_1$ = $-$0.79 $\pm$ 0.09 ($\chi^2/\nu=1.09$ for
$\nu=7)$; $\alpha_2$ = $-$0.36 $\pm$ 0.05 ($\chi^2/\nu=1.23$ for
$\nu=9)$ for the $B$ band and $\alpha_2$ = $-$0.52 $\pm$ 0.15
($\chi^2/\nu=7.7$ for $\nu=9)$ for the $I$ band. The best-fit
parameters for the $B$, $V$, $R$, and $I$ bands are summarized in
Table 2. Our results show not only the clear presence of mild breaks
in the $V$, $R$ band lightcurves but a flattening trend after the
break. Furthermore, our $R$ band slope before the break ($\alpha \sim
-0.84$) is in agreement with the corresponding value derived by Quimby
et al. (2006) for the interval between 0.0019 and 0.05 days after the
burst.

\subsection{Color and spectral flux distribution}

Our multi-wavelength observations indicate that median colors between
0.07 and 0.35 days are $V-R = 0.45 \pm 0.11$, $R-I = 0.46 \pm 0.10 $,
and $B-V = 0.84 \pm 0.14 $. These values have been corrected for
foreground reddening of E($B-V) = $ 0.011 mag (Schlegel et
al. 1998). The $V-R$ and $R-I$ colors so derived are consistent with
those of the typical long GRBs (Simon et al. 2001), but the $B-V$
color is slightly redder than those of the typical long GRBs ($B-V =
0.47 \pm 0.17$). The larger $B-V$ value may imply a certain absorption
effect because the redshift of GRB\,050319 was determined to be 3.24
(Fynbo et al. 2005).

The $B$, $V$, $R$ and $I$ magnitudes have been further converted to
fluxes using the effective wavelengths and normalizations of Fukugita
et al. (1995). The effect of the Galactic interstellar extinction has
been corrected. Figure 2 shows two samples of spectral energy
distribution obtained by LOT at 0.13 and 0.21 days after the
occurrence of GRB\,050319. A drop in the $B$ band flux at about
$4380\AA$ can be clearly seen. We subsequently fitted the flux
distribution of $V$, $R$ and $I$ bands with a power-law function $
F(\nu,t) \propto \nu^{\beta}$; here $F(\nu,t)$ is the flux at
frequency $\nu$ with a certain $t$ and $\beta$ is the spectral
index. We find that $\beta = -1.08 \pm 0.05$ ($\chi^{2}/\nu$= 0.05 for
$\nu=1$) at 0.13 days and that $\beta = -1.08 \pm 0.32$
($\chi^{2}/\nu$ = 2.3 for $\nu=1$) at 0.21 days. Our result ($\beta =
-1.08$ with a rms error 0.23) is consistent with the X-ray fitting
value ($\beta = -0.69 \pm 0.06$) in a 3-$\sigma$ level.

With a redshift of 3.24, the Ly${\alpha}$ absorption feature would
shift into the $B$ bandpass, causing reduction of the afterglow flux
in the $B$ band.  To correct for this absorption effect, we used the
formulation derived by Yoshii et al. (1994), in which the optical
depth is a function of the observed wavelength and source
redshift. With the computed optical depth in the $B$ band, and a
spectral slope of $\beta = -1.08$, we found the expected $B$ band
magnitude after Ly${\alpha}$ absorption to be $21.33 \pm 0.05$ at 0.13
days. This value compares very well with our observed value of $B$ =
21.26 $\pm$ 0.17 at 0.13 days after correction for Galactic
extinction. The drop at the B band is hence fully produced by the
Ly${\alpha}$ absorption and no spectral breaks should have taken place
during our observation.
 
\section{Discussion and Summary}

It is important to note that {\it Swift} found two breaks at 0.004 and
0.313~days after the burst in the X-ray afterglow observations
(Cusumano et al.  2006), but we only found a single break in our $V$
and $R$ lightcurves (see Figure 1). In the following since there are
more data points available for the $R$ band data we will focus on
this. It is useful to remember that $\alpha_1$ = $-0.84$ and
$\alpha_2=-0.48$ at the break time of $t_{\rm b}$ = 0.04~days.

\subsection{Before the  optical break ( t $<$ 0.04 days)}

The slope $\alpha_1$ (=$-$0.84) is consistent with the typical range
of $\alpha = -0.62$ to $-2.3$ for many well observed GRBs. According
to the standard afterglow model relating the power-law index
($\alpha$) to the power-law index ($\it{p}$) of an electron spectrum
(Zhang \& M\'esz\'aros, 2004; Dai \& Cheng, 2001), the corresponding
value for $\alpha_1=-0.84$ is $\it{p}$ = 2.1, which is in agreement
with the constant-density ISM model with slow cooling in which $\it{p}
>$ 2 for $\nu_{\rm m}< \nu_{\rm opt} < \nu_{\rm c}$ ($\nu_{\rm m}$ is
the typical frequency; $\nu_{\rm opt}$ is the optical frequency;
$\nu_{\rm c}$ is the cooling frequency). In light of the XRT
observations, the first break was likely caused by the transition from
the tail end of the low energy prompt emission to the afterglow phase
(Zhang et al. 2006). However, it is important to note that the X-ray
break at 0.004 days (where the steeper slope becomes shallow) is not
accompanied by an $R$ band break. At the same time, the power-law
decay slope in the X-ray ($\sim -$ 5.53) and the $R$ band ($\sim -$
0.84) are quite different. This is an indication not only that the
behaviors of the X-ray afterglow and optical afterglow of the
GRB\,050319 event are different, but also suggested that the afterglow
phase already dominated the optical bands when the optical emission
was first detected.

\subsection{Shallow Decay}

The power law index becomes shallow after the break ($t_{\rm b}$ =
0.04 days). Neither the jet (Rhoads 1999) nor the break frequencies
across the optical wavelength (Sari et al., 1998) suitably explain the
break they see in the GRB\,050319 lightcurves. As discussed before,
the X-ray lightcurve between the two breaks 0.004 and 0.313~days after
the burst is also characterized by shallow decay. Zhang et al. (2006)
suggested that such behavior is related to continuous energy injection
into the ISM. For a long-lasting central engine, the energy injection
rate is $\dot{E}(t) \propto t^{-q}$ and with $q < 1$ (Zhang \&
M\'esz\'aros 2001). For slow cooling in the ISM, the temporal index
can be expressed as : $\alpha = [(2p-6)+(p+3)q] / 4 = [(q-1) +
  {(2+q)\beta] / 2},$\, when\, $\nu_{\rm m} < \nu < \nu_{\rm
  c}$. Using this formulation, Zhang et al. (2006) obtained $q$ = 0.6
and $p$ = 2.4 from the X-ray observations. With $\alpha$ = $-$0.48 and
$\beta$ = $-$1.08 from the $R$ band observations, we find that $q$ =
0.72 and $p$ = 2.12. The results not only indicate that the electron
spectrum power law index is the same before and after the break, they
also compare well with the results of Zhang et al. (2006). These
results indicate that the shallow decays evidenced by both X-ray and
optical afterglows could be of a similar origin, related to a
continuous energy injection mechanism.

According to the energy injection model, would also expect an X-ray
break at the time of the optical flattening break, because the onset
of the energy injection should also alter the X-ray temporal
index. However, such an X-ray break is not observed, which suggests
that some modifications to the injection model may be needed. As
mentioned in $\S$ 4.1, the X-ray break at 0.004 days was not
accompanied by a break in the $R$ lightcurve. Although a chromatic
break in the X-ray was found at 0.004 days and in the optical region
at 0.04 days, the lightcurves at both wavelengths indeed showed
shallow decay after the breaks, which can be explained by the energy
injection model. However, it is difficult for energy injection from
0.004 days to 0.04 days to affect only high energies. This difficulty
indicates that energy injection is an imperfect mechanism for
explaining the shallow optical or X-ray phase associated with the
GRB\,050319 event.

Several models have recently been proposed to explain the shallow
decay effect. Using the multiple-subjet model (Nakamura, 2000), Toma
et al. (2006) invoked the superposition of afterglows from many
off-axis subjets. Eichler \& Granot (2006) favored a combination of
the tail of prompt emission model with the afterglow emissions
observed from a viewing angle outside the edge of the jet. These
arguments hence suggest that the multiple-subjet model and the
patchy-shell model (Kumar \& Piran, 2000) might provide a theoretical
basis for explaining the observed shallow decays in the X-ray and
optical lightcurves. It is interesting to note that in order to
sustain the shallow decay process these models all require high
gamma-ray efficiency (75-90$\%$); additional mechanisms such as prior
activity (Ioka et al. 2005) and time-dependent shock generation (Fan
\& Piran, 2006) have been also proposed. Comprehensive
multi-wavelength observations, such as those reported here, provide
important keys to improve these models.

Finally, the second break in the X-ray emissions ($\sim$ 0.313~days
after the burst), has been interpreted as being due to an unusual flat
jet break (Cusumano et al. 2006), Zhang et al. (2006) however provided
an alternate explanation, a sudden cessation of the energy
injection. In both interpretations, the corresponding break should
appear in both the X-ray and optical lightcurves. This effect cannot
be clearly identified in our measurements until 0.413 days after the
burst. The lack of data for the subsequent time interval could lead to
uncertainty in the power-law fitting. We thus cannot fully exclude the
existence of a second break in the optical lightcurves. However,
Panaitescu et al. (2006) have studied several afterglows. They found
the shallow power-law decay evidenced by the X-ray emissions to
steepen about 0.04$-$0.17 days after the burst, although there was no
accompanying break found in the optical range. They suggest that such
chromatic X-ray breaks may be common. The chromatic breaks (e.g. the
shallow X-ray phase becomes steeper, with no accompanying optical
break) may be caused by differences in the X-ray and optical outflow
(Panaitescu et al. 2006), or by changes in the typical electron energy
parameters (the so-called microphysical parameters) at the end of
energy injection (Panaitescu 2006).

In summary, our analysis of the optical multi-wavelength observations
of GRB\,050319 made at Kiso and Lulin compared with the X-ray
observations from {\it Swift} found the following major results:

\begin{itemize}

\item The $B$, $V$, $R$, and $I$ band lightcurves displayed unusual
shallow decays.

\item The $R$ lightcurve can be described by a smooth broken power-law
function. $\alpha_1 \sim -0.84$ becomes shallow ($\alpha_2 \sim
-0.48$) 0.04 days after the occurrence of the GRB.

\item The shallow decay observed in the X-ray and optical lightcurves
may have a similar origin related to energy injection. However, our
observations indicate that there is still a major puzzle remaining
concerning the chromatic breaks in the $R$ band lightcurve (at
0.04~days) and the X-ray lightcurve (at 0.004~days).

\item Our calculations revealed that the drop in spectral energy
distribution was fully caused by a shift in the Ly${\alpha}$
absorption to the $B$ bandpass at $z$ = 3.24.

\end{itemize}

\acknowledgments 

We thank the referee for his/her valuable advice. This work is
supported by NSC\,95-2752-M-008-001-PAE, NSC\,95-2112-M-008-021, the
Japan Society for the Promotion of Science (JSPS) Grant-in-Aid for
Young Scientists (B) No.\,18740147 and JSPS Research Fellowships for
Young Scientists (Y.U.).

\clearpage 

\begin{table}
\begin{center}
\caption{GRB\,050319 Optical Afterglow Photometry\tablenotemark{a}}
\begin{tabular}{cccc}
\tableline \tableline 
Days after GRB\tablenotemark{b}  &  Filter & Magnitude\tablenotemark{c} & Site \\
\tableline
  0.05443 &    R &   19.01$\pm$0.08 & Kiso\\
  0.06023 &    B &   20.42$\pm$0.13 & Kiso\\
  0.08037 &    V &   19.75$\pm$0.07 & Lulin \\
  0.10439 &    I &   19.09$\pm$0.07 & Lulin\\
  0.27821 &    B &   20.92$\pm$0.12 & Kiso\\
  0.33424 &    I &   19.91$\pm$0.09 & Lulin\\
  0.40097 &    V &   20.68$\pm$0.12 & Lulin\\
  0.41346 &    R &   20.20$\pm$0.07 & Lulin\\
\tableline
\tablenotetext{a}{Table 1 is published in its entirety in the electronic edition of the $\it{Astrophysical\,Journal}$.}
\tablenotetext{b}{The burst time is 2005 March 19, UT 09:29:01.44}
\tablenotetext{c}{The magnitudes are not corrected for Galactic extinction.}
\end{tabular}
\end{center}
\end{table}

\clearpage

\begin{table}
\begin{center}
\caption{Fitting results of the GRB\,050319 lightcurves : The $B$ and
$I$ band data were fitted by a respective power-law model [$F \propto
t^{\alpha}$] before and after the break ($t_b =0.04\,{\rm days}$). On
the other hand, the $V$ and $R$ data were fitted by a smoothly broken
power-law [equation (1)].}
\begin{tabular}{cccc}
\tableline \tableline 
 Filter    &  $\alpha_1$ & $\alpha_2$        &  $t_b ($\rm day$)$\\ 
\tableline
  B        & -0.79$\pm$0.09 & -0.36$\pm$0.05 &    -\\
  V        & -0.87$\pm$0.21 & -0.49$\pm$0.05 & 0.042$\pm$0.058\\
  R        & -0.84$\pm$0.02 & -0.48$\pm$0.03 & 0.046$\pm$0.008\\
  I        &       -        & -0.52$\pm$0.15 &     -\\  
\tableline

\end{tabular}
\end{center}
\end{table}

\clearpage

\begin{figure}
\begin{center}
\includegraphics[width=0.8\textwidth]{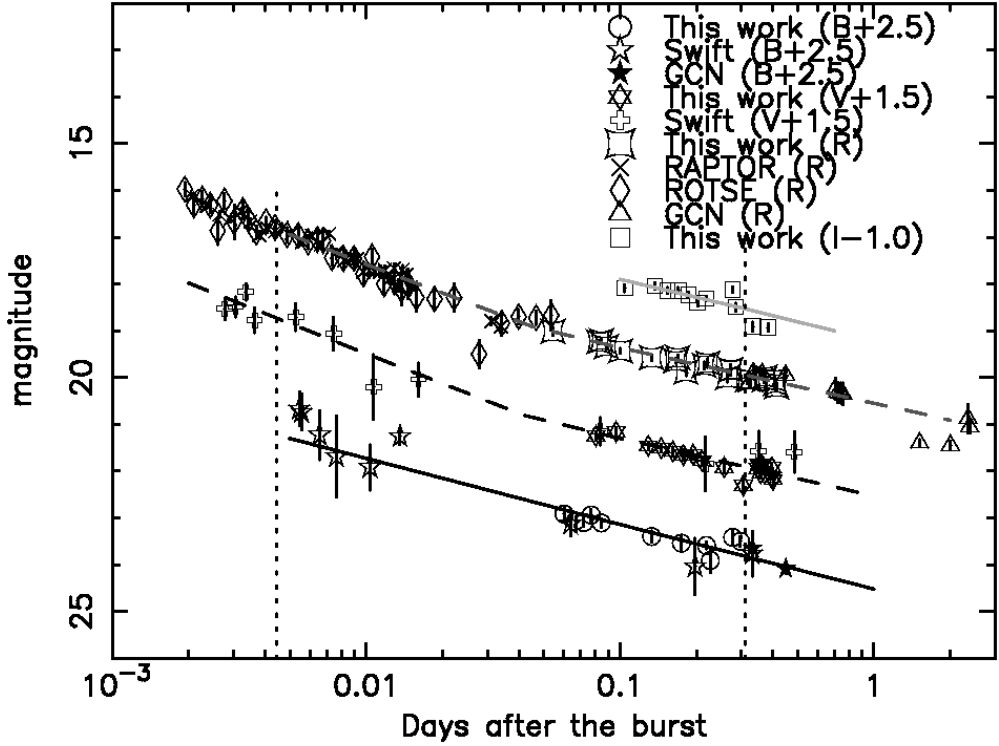}
\caption{Optical lightcurves of GRB\,050319 : The solid lines present
the best fit by the single power-law model [$F \propto t^{\alpha}$]
for the $B$ band ($\alpha = -0.56 \pm 0.06$) and the $I$ band ($\alpha
= -0.52 \pm 0.15$). The dashed line indicates the best fit by the
smooth broken power-law [equation (1)] with the $V$ band ($\alpha_1 =
-0.87 \pm 0.21$, $\alpha_2 = -0.49 \pm 0.05$, $t_{\rm b}$ = 0.042
$\pm$ 0.058 days) and the $R$ band ($\alpha_1 = -0.84 \pm 0.02$,
$\alpha_2 = -0.48 \pm 0.03$, $t_{\rm b}$ = 0.046 $\pm$ 0.008
days). These observations are based on our EAFON data and Quimby et
al. (2006), Wo\'zniak et al. (2005), Mason et al. (2005), Greco et
al. (2005), Kiziloglu et al. (2005), Misra et al. (2005), and Sharapov
et al. (2005a,b). The dotted lines represent the break times of X-ray
afterglows at 0.004 and 0.31 days after the burst. The two breaks were
not found in X-ray observations, but a mild break seems to exist in
the $V$ and $R$ band lightcurves. }
\end{center}
\end{figure}

\clearpage

\begin{figure}
\begin{center}
\includegraphics[width=0.8\textwidth]{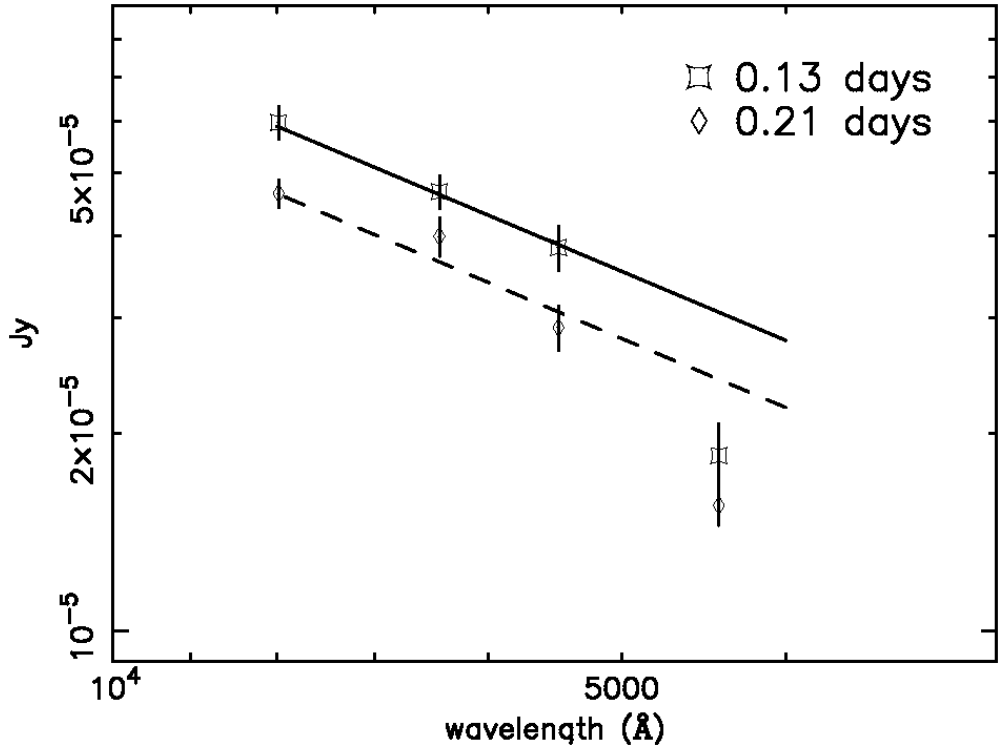}
\caption{Spectral energy distribution of GRB\,050319 between the $V$,
$R$ and $I$ bands at 0.13 and 0.21 days after the burst (corrected for
Galactic extinction). The solid and dashed lines indicate the best fit
by the power-law model [$F(\nu) \propto \nu^{\beta}$], where $\beta =
-1.08 \pm 0.05$ for 0.13 days and $\beta = -1.08 \pm 0.32$ for 0.21
days, respectively.}
\end{center}
\end{figure}


\clearpage


\begin{thebibliography}{}

\bibitem[Blustin (2006)]{blustin} Blustin, E. et al. 2006 \apj, 637, 901
\bibitem[Cusumano (2006)]{cusumano} Cusumano, G. et al. 2006, \apj, 639, 316
\bibitem[Dai \& Cheng (2001)]{dai2} Dai, Z. G., \& Cheng, K. S. 2001, \apj, 558, 109
\bibitem[Eichler et al.(2006)]{eichler} Eichler, D., \& Granot, J. 2006, \apj, 641, L5
\bibitem[Fan (2006)]{fan} Fan, Y., \& Piran, T. 2006, \mnras, 369, 197
\bibitem[Fynbo (2005)]{fynbo} Fynbo, J. P. U. et al. 2005, GCN Circ. 3136
\bibitem[Fukugita 1995]{fukugita} Fukugita, M., Shimasaku, K., \& Ichikawa, T. 1995, PASP, 107, 945
\bibitem[Gehrels et al. (2004)]{gehrels} Gehrels, N.  et al. 2004 \apj, 611, 1005
\bibitem[George et al. (2006)]{George} George, K., Banerjee, D. P. K., Chandrasekhar, T., \& Ashok, N. M. 2006 \apj, 640, L13 
\bibitem[Greco et al. (2005)]{greco} Greco, G., Bartolini, C., Guarnieri, A., Piccioni,A., Ferrero, P. \& Bruni, I. 2005, GCN Circ. 3142
\bibitem[Henden (2005)]{Henden}Henden, A. 2005, GCN Circ. 3454
\bibitem[Ioka (2005)]{Ioka} Ioka, K., Toma, K., Yamazaki, R., \& Nakamura, T. 2005, A\&A in press (astro-ph/0511749)
\bibitem[Kinoshita et al.(2005)]{kinoshita} Kinoshita, D., et al. 2005, ChJAA, 5, 315
\bibitem[Kiziloglu et al. (2005)]{kiziloglu et al.} Kiziloglu, U. et al. 2005, GCN Circ. 3139
\bibitem[Krimm et al. (2005)]{Krimm05a} Krimm, H. et al. 2005, GCN Circ. 3119
\bibitem[Kumar 2000]{kumar} Kumar, P. \& Piran, T.  2000, \apj, 535, 152
\bibitem[Mason (2006)]{mason} Mason, K. O. et al. 2006, \apj, 639, 311
\bibitem[Misra et al. (2005)]{misra} Misra, K., Kamble, A. P. \& Pandey, S.B. 2005, GCN Circ. 3130
\bibitem[Nakamura 2000]{nakamura} Nakamura, T.  2000, \apj, 534, L159
\bibitem[Nousek (2006)]{nousek} Nousek, J. A. et al. 2006, \apj, 642, 389
\bibitem[Panaitescu et al. (2006)]{panaitescu} Panaitescu, A., M\'esz\'aros, P., Burrows, D., Nousek, J., O'Brien, P. \& Willingale, R. 2006, MNRAS, 369, 2059
\bibitem[Panaitescu (2006)]{panaitescu2} Panaitescu, A., 2006, preprint (astro-ph/0607396)
\bibitem[Quimby et al. (2006)]{quimby} Quimby, R. M.  et al. 2006, \apj, 640, 402
\bibitem[Rhoads et al. (1999)]{quimby} Rhoads, J. E. 1999, \apj, 525, 737

\bibitem[Rykoff et al.(2005)]{rykoff} Rykoff, E., Schaefer, B., \& Quimby, R. 2005, GCN Cir. 3116
\bibitem[Rykoff et al.(2006)]{rykoff06} Rykoff, E. et al. 2006, \apj, 638, L5
\bibitem[Sari (2005)]{sari} Sari, R., Piran, T., \& Narayan, R. 1998, \apj, 497, L17
\bibitem[Schlegel, Finkbeiner, \& Davis (1998)]{schlegel} Schlegel, D. J., Finkbeiner, D. P., \& Davis, M. 1998, ApJ, 500, 525
\bibitem[Sharapov et al. (2005a)]{sharapov05a} Sharapov, D. et al. 2005a, GCN Circ. 3124
\bibitem[Sharapov et al. (2005b)]{sharapov05b} Sharapov, D. et al. 2005b, GCN Circ. 3140
\bibitem[Simon (2001)]{simon} Simon, V., Hudec, R., Pizzichini, G., \& Masetti, N. 2001, A\&A, 377, 450
\bibitem[Stetson (1987)]{Stetson87} Stetson, P. B. 1987, PASP, 99, 191
\bibitem[Toma et al.(2006)]{toma} Toma, K., Ioka, K., Yamazaki, R. \& Nakamura, T. 2006, \apj 640, L139
\bibitem[Torii et al. (2005)]{torii} Torii, K. 2005, GCN Circ. 3121
\bibitem[Urata et al. (2005)]{urata} Urata, Y. et al. 2005, Nuovo Cimento, 28, 775
\bibitem[Wo\'zniak et al.(2005)]{wozniak} Wo\'zniak, P. R., Vestrand, W. T., Wren, J. A., White, R. R., Evans, S. M., \& Casperson, D. 2005, ApJ, 627, L13
\bibitem[Yoshii(1994)]{yoshii} Yoshii, Y. \& Peterson, B. A. 1994, \apj, 436, 551
\bibitem[Yoshioka et al. (2005)]{yoshioka} Yoshioka, T. et al. 2005, GCN Circ. 3120
\bibitem[Zhang \& M\'esz\'aros (2001) ]{zhang3} Zhang, B., \& M\'esz\'aros, P. 2001, \apj, 552, L35
\bibitem[Zhang \& M\'esz\'aros (2004) ]{zhang1} Zhang, B., \& M\'esz\'aros, P. 2004, Int. J. Mod. Phys. A., 19, 2385
\bibitem[Zhang (2006) ]{zhang2} Zhang, B. et al. 2006, \apj, 642, 354
\end{thebibliography}
\end{document}